# Micrometer-scale displacement and thickness sensing using a single terahertz resonant-tunneling diode

*Li Yi[1,2], Shota Ito[1], Chao Tang[3], Yousuke Nishida[4], Koji Terumoto[4], Toshihisa Maeda[4], Yuta Inose[2], and Masayuki Fujita[2]

1 Graduate School of Science and Engineering, Ibaraki University, Hitachi, Japan

2 Graduate School of Engineering Science, The University of Osaka, Toyonaka, Japan

3 Frontier Research Institute for Interdisciplinary Sciences, Tohoku University, Sendai, Japan

4 ROHM Research & Development Center, ROHM Co., Ltd., Kyoto, Japan

*Corresponding author: li.yi.wg60@vc.ibaraki.ac.jp

## Abstract

Resonant tunneling diodes (RTDs) support room-temperature terahertz (THz) oscillation and simultaneous THz-band detection, enabling compact monostatic THz sensors for practical and cost-effective sensing applications. In this paper, we present a highly integrated 280 GHz-band radar system based on a single RTD that exploits the self-mixing effect to generate a low-frequency interferometric signal. The resulting self-mixing signal is further analyzed from a radar perspective and processed to extract micrometer-scale displacement and thin-film thickness variations. Experimentally, the proposed system demonstrates a minimum detectable displacement of approximately 5 μm and quantitatively resolves polymer film thicknesses of 12.5, 25, and 50 μm.

Contributions:

- 1) Single-device THz transceiver: A single RTD is operated as both a bias-tunable oscillator and a self-oscillating mixer via self-mixing.
- 2) Low-frequency control/readout: The sensing signal is confined to sub-MHz bandwidth, enabling low-cost MHz-rate acquisition electronics.
- 3) Interferometric processing: A phase-based processing approach and a simple phase estimator are introduced for micrometer-order displacement/thickness sensing.

## I. INTRODUCTION

Terahertz (THz) wave (~0.1–10 THz) provides a unique combination of sub-millimeter-scale wavelength, material-dependent dielectric response, and the ability to penetrate many common nonconducting materials. These properties motivate a wide range of nondestructive evaluation, imaging, and metrology applications [1-5]. Among established techniques, terahertz time-domain spectroscopy (THz-TDS) provides broadband permittivity extraction and time-of-flight-based thickness/distance estimation [3-4], but its reliance on femtosecond lasers and complex optoelectronics limits system compactness and cost efficiency.

As an alternative, continuous-wave (CW) radar approaches offer an attractive alternative when compactness and low power are prioritized [2,5-9]. The CW radar technique is commonly implemented using amplitude- or frequency-modulated continuous-wave (AMCW/FMCW) techniques, which translate target distance into a measurable phase delay term [6]. These approaches eliminate ultrafast optical components and enable phase and spectrum processing in THz-band systems, with many successful applications have already been reported [2,5,10-13]. However, they typically rely on high-end THz hardware, including a tunable high-frequency source as well as dedicated transmit, receive, and mixing chains. Although CMOS/SiGe-based solutions have become increasingly practical toward the sub-THz band, efficient implementations above 300 GHz remain challenging in terms of output power, tuning range, and front-end loss [11-12].

To further improve system integration and reduce cost at terahertz frequencies, resonant tunneling diodes (RTDs) have emerged as attractive candidates owing to their ability to sustain room-temperature THz oscillation together with strong intrinsic nonlinearity [13–15]. Numerous sensing applications have demonstrated their remarkable sensitivity and high level of integrability [18-21]. Notably, RTDs can also operate under injection-locking conditions [16,17], a phenomenon extensively studied in laser technology [22-24]. This mechanism has been shown to enhance the sensitivity of THz-band detectors [17] and suggests additional opportunities for sensing applications. Recently, we demonstrated THz-band three-dimensional radar imaging using a single RTD operating in a self-mixing configuration, in which the signal reflected from the target is reinjected into the same RTD [18]. This approach highlights significant advantages in system size, cost, and power consumption compared with conventional THz radar architectures.

In this work, we leverage RTD self-mixing in combination with an interferometric radar technique to realize micrometer-scale displacement and thin-film thickness sensing around 280 GHz using a single RTD. We (i) present an interferometric signal model tailored to RTD self-mixing, (ii) introduce a simple phase-shift estimator based on this model, and (iii) quantify the robustness of the proposed method through both simulation and experimental results, and (iv) validate the approach experimentally for 5–200 μm displacement and 12.5–50 μm polymer film thickness steps. Compared with conventional THz radar architectures, the proposed approach significantly simplifies the front-end hardware while retaining coherent phase sensitivity, offering a practical and scalable solution for THz-band sensing applications.

## II. METHODOLOGY

### A. THz sensing using RTD self-mixing signal

A resonant-tunneling diode (RTD) is a quantum heterostructure device that exhibits negative differential conductance (NDC) in its current–voltage characteristic owing to resonant tunneling [13]. When the magnitude of the NDC exceeds the loss of an external resonant circuit, self-sustained oscillation can be generated at sub-terahertz or terahertz frequencies [14]. Owing to the strong nonlinearity of the current–voltage (I–V) characteristic, the same device can operate as both a terahertz source and a detector, as illustrated in Fig. 1. Moreover, within a specific bias range the RTD can sustain oscillation while receiving a reflected signal, enabling self-oscillating mixing; this operating regime is commonly referred to as the self-mixing region.

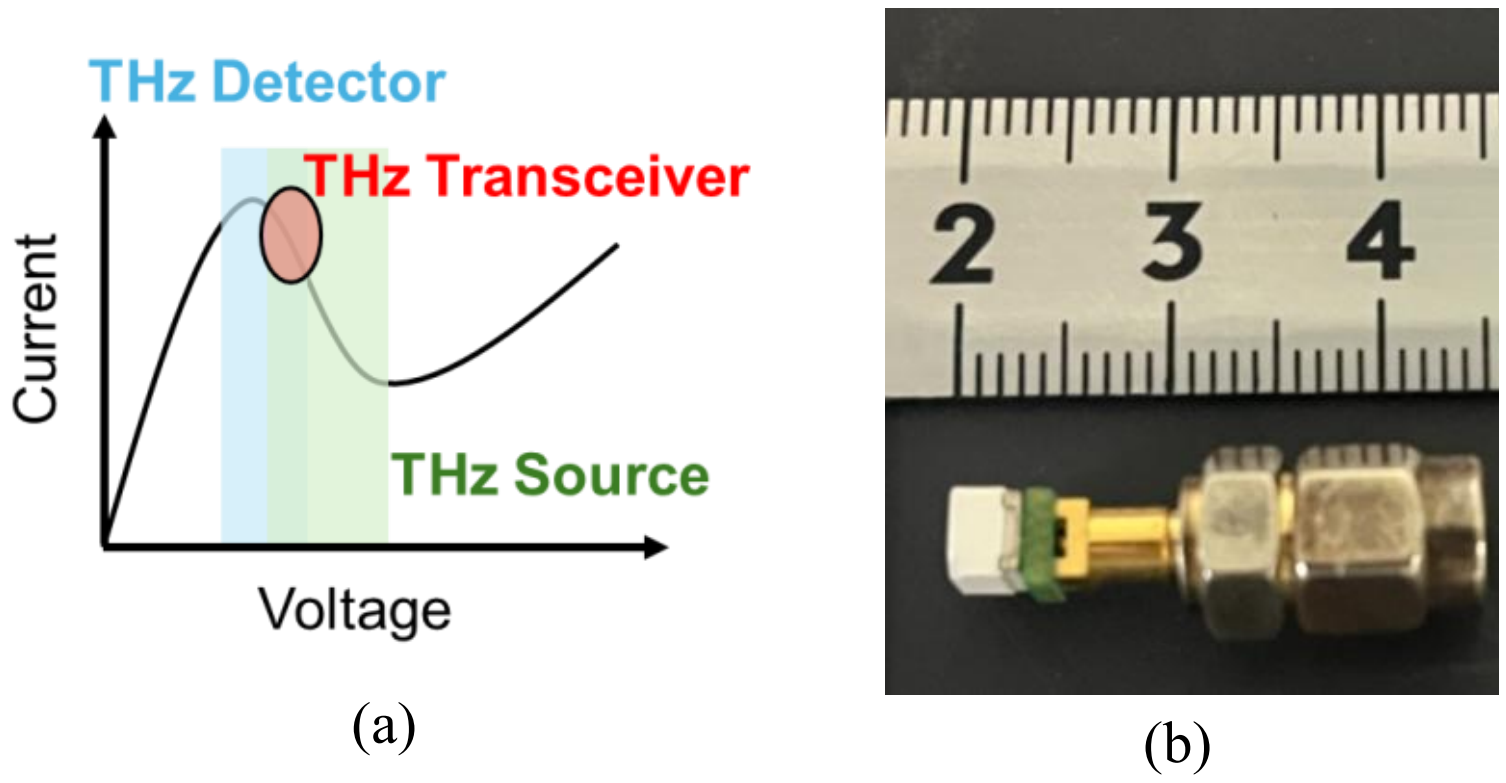

Fig. 1. (a) Conceptual RTD I–V characteristic and bias regions used for oscillation and self-oscillating mixing; (b) photograph of the RTD module used in this work.

For the sensing application, if a portion of the radiated THz signal is reflected by the target and reinjected into the RTD oscillator, giving rise to a low-frequency baseband response commonly referred to as self-mixing. This self-mixing effect enables a single RTD to simultaneously function as a signal source and a nonlinear mixer, thereby supporting highly compact transceiver architectures that consolidate oscillation, mixing, and detection within a single device [17,18]. Notably, the self-mixing effect in an RTD is also referred to as external feedback and is formally analogous to optical self-mixing in semiconductor lasers [16,22].

In both cases, the oscillator dynamics can be modeled as a delay-feedback system closely related to the Lang–Kobayashi framework [24], in which the feedback strength and phase govern waveform distortion and phase sensitivity. According to the Lang–Kobayashi framework, $\phi_0(f)$ denotes the free-running phase in the absence of self-mixing. Following the standard excess-phase relation used in self-mixing interferometry [22], the free-running phase $\phi_0(f)$ and the feedback-modified phase $\phi(f)$ satisfy

$$\phi_0(f) = \phi(f) + C\sin\left(\phi(f) + \psi\right), \qquad (1)$$

and the round-trip propagation phase is:

$$\phi_0(f) = \frac{4\pi f R}{c}. \qquad (2)$$

Here, $c$ is the speed of light, $f$ is the operating frequency set by the RTD bias voltage, $R$ is the target range, $C$ is the feedback coupling factor, and $\psi$ is a constant phase offset. This indicates that when the emitted field is reflected by a distant target and a portion couples back to the RTD antenna, producing a bias-dependent self-mixing voltage that encodes the round-trip phase. The measurable self-mixing signal can be described as a phase-dependent interferometric response. In optical feedback self-mixing interferometry, the detected signal is commonly expressed as a cosine function of the phase [22,23]. Accordingly, we adopt the simplified form

$$S(f) = \cos(\phi(f)), \qquad (3)$$

where $\phi(f)$ is the feedback-modified phase determined from Eq. (1). This cosine representation is consistent with the standard weak-feedback description of self-mixing interferometry, as illustrated in Fig. 2(a). Although Eq. (3) has a cosine form, the resulting waveform can deviate significantly from a pure sinusoid because $\phi(f)$ is generally a nonlinear function of frequency under external feedback. As the feedback strength $C$ increases, the waveform becomes increasingly distorted, while remaining periodic, indicating that the range information is preserved for accurate distance estimation [18,22], as shown in Fig. 2(b). Moreover, when $C$ is close to unity, the sharpened waveform provides multiple localized high-contrast features across the sweep, making global waveform alignment more sensitive to small phase translations than a pure sinusoid, where displacement information is concentrated near quadrature points.

Notably, additional attenuation may be required to avoid the strong-feedback regime ($C > 1$), in which the implicit phase equation can admit multiple solutions, leading to branch switching, hysteresis, and abrupt fringe discontinuities that complicate phase demodulation and degrade robust delay estimation [22,23]. In this work, we restrict attention to the $0 < C < 1$ regime, consistent with prior reports and our experimental observations using a spatially separated or collimated beam [18], whereas the detailed validation is discussed further in Sec. III(B).

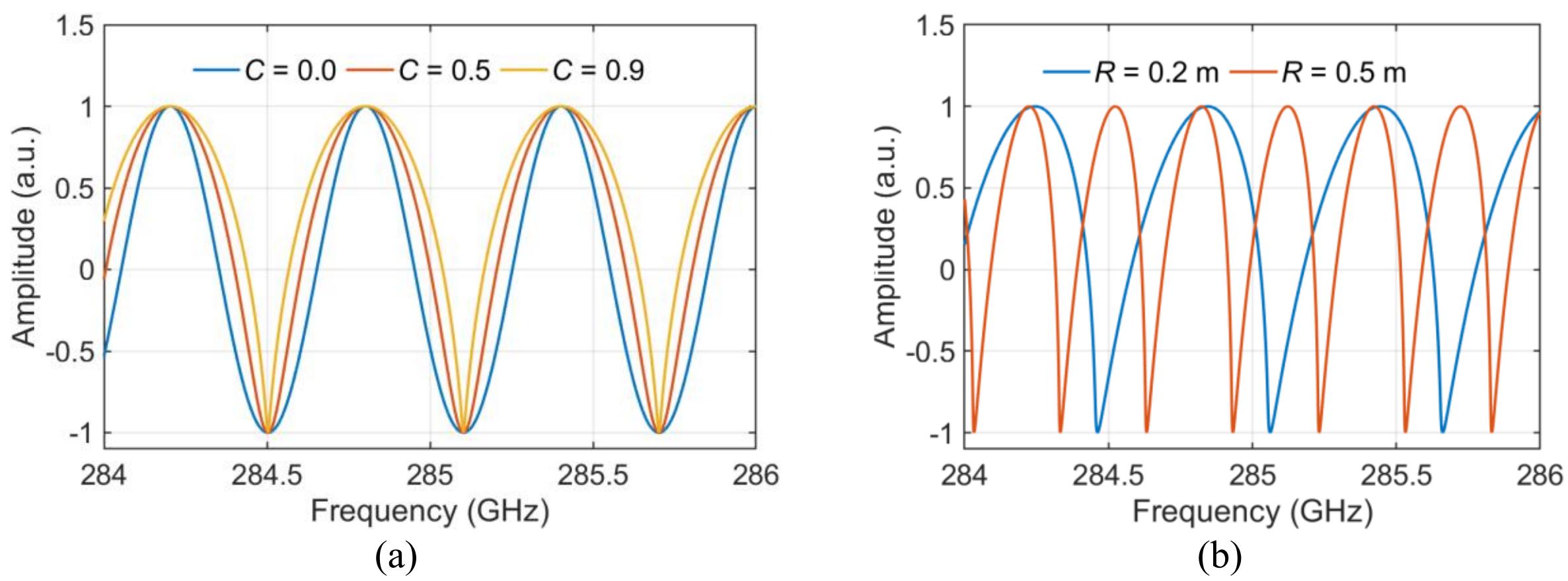

**Fig. 2.** Simulated RTD self-mixing signals over a 2-GHz bandwidth at 280 GHz: (a) signals for different feedback coupling factors $C$(with $R = 0.2$ m and $\psi = 0$); (b) signals for different target ranges $R = 0.2$ m and 0.5 m (with $C = 0.9$ and $\psi = 0.5$).

Building on the self-mixing capability demonstrated in Fig. 2(b), a single RTD can be employed to estimate the target distance in a manner analogous to a conventional radar system. Compared with the conventional CW radar architecture illustrated in Fig. 3, the proposed approach integrates oscillation and mixing within a single device. However, radar sensing based on the self-mixing signal differs from a conventional radar system in several key aspects, which are discussed in the next section, and the implementation details of the proposed system are further described in Sec. III(A).

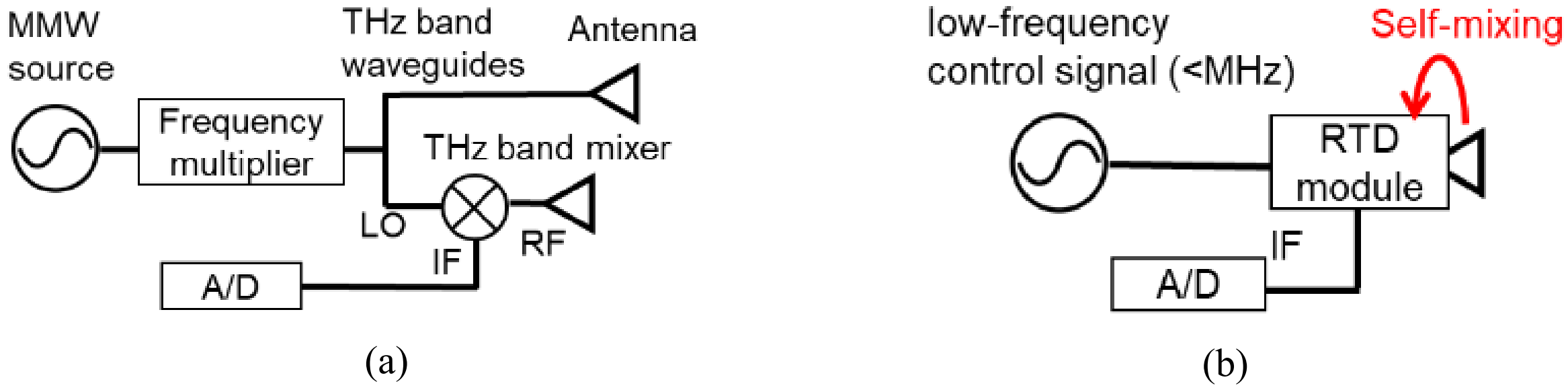

Fig. 3. (a) Conventional radar front-end with separate transmitter and receiver; (b) proposed RTD self-mixing sensing configuration.

## B. Comparison of self-mixing signal and radar beat signal

In a standard CW radar configuration, a single-tone signal at frequency $f$ is transmitted toward a target at range $R$ and coherently mixed with a reference derived from the transmitter, as shown in Fig. 3(a) [6]. The detected coherent signal can be written as [7]

$$S_1(f) = A(f)\exp\left(j(\phi_0(f) + \phi_n)\right), \qquad (4)$$

where the round-trip propagation phase $\phi_0$ is given in Eq. (2), and $\phi_n$ denotes phase noise and other frequency-independent phase terms introduced by the measurement chain. In what follows, we omit the amplitude term $A(f)$ and the constant phase offset $\phi_n$ for notational convenience.

To extract range information embedded in the phase term, the FMCW radar concept employs a swept carrier frequency. The frequency dependence of $\phi_0(f)$ encodes the target distance as a low-frequency beat signal, from which the range can be estimated using a Fourier transform. [2, 6]. Notably, taking the real part of the coherent beat signal yields a sinusoidal waveform that is equivalent to the self-mixing signal in the limiting case $C = 0$. In other words, as the injected feedback power becomes smaller, the RTD self-mixing signal approaches the conventional radar beat signal.

Compared with conventional radar architectures that rely on a tunable source, low-loss waveguide/quasi-optical front ends, and coherent mixer/LO chains, the proposed RTD self-mixing setup can be significantly more compact and cost-effective, as shown in Fig. 3. However, the feedback coupling factor is inherently target dependent, since it is determined by the power coupled back into the RTD, as described by Eq. (1). As a result, the applicability of the RTD approach is mainly limited to scenarios with stable and well-defined return paths. In addition, operation in the self-mixing regime typically reduces the effective tuning range of the RTD, making it challenging to achieve high-resolution range profiling. In conventional radar, the range resolution is approximately

$$\Delta R = \frac{c}{2B}, \qquad (5)$$

where $B$ is the chirp bandwidth. In contrast to THz-TDS, which employs ultrashort pulses with an effective bandwidth of several terahertz and thereby enables micrometer-scale range resolution for thin-film thickness measurements [3, 9], the maximum tuning bandwidth of an RTD is typically limited to ~10% of its center frequency. Consequently, in the 280 GHz band, the achievable range resolution is on the order of centimeters.

## C. Simplified interferometric radar technique using self-mixing signal

Given the characteristics of the proposed RTD self-mixing system, it is more advantageous to exploit its high carrier frequency and compact implementation than to depend on an extremely large bandwidth. To this end, super-resolution techniques based on differential phase information have been widely adopted in laser interferometric measurement and radar remote sensing [7], [8], [22]. In such an interferometric framework, comparison between a displaced signal and a reference signal enables the estimation of displacements far smaller than the wavelength from small phase perturbations. The underlying principle is that high resolution can be achieved through a high carrier frequency rather than a large bandwidth, as indicated by (5), making this approach particularly suitable for RTD self-mixing signals.

For a conventional radar configuration, let $S_1(f)$ and $S_2(f)$ denote the reference and displaced signals, respectively. When a small displacement $\Delta d$ changes the target range from $R$ to $R + \Delta d$, the resulting complex beat signal $S_2(f)$ can be written as

$$S_2(f) = A(f)\exp\left(-j\frac{4\pi f}{c}(R + \Delta d)\right) = S_1(f)\exp\left(-j\frac{4\pi f}{c}\Delta d\right). \qquad (6)$$

Accordingly, the displacement introduces an additional propagation phase that varies linearly with frequency, and the corresponding phase difference can be written as [8]

$$\Delta\phi_0(f) \triangleq \angle\{S_2^*(f)S_1(f)\} = \frac{4\pi f}{c}\Delta d. \qquad (7)$$

Here, $(\cdot)^*$denotes complex conjugation, and $\angle\{\cdot\}$extracts the relative phase. Equation (7) shows that a small displacement appears as a perturbation in the free-running propagation phase. However, this phase perturbation is not strictly constant over the bandwidth. To approximate it as frequency-independent across a bandwidth $B$, the narrowband condition requires [8]

$$\Delta\phi_{\text{across}} = \frac{4\pi B}{c}\Delta d \ll 1. \qquad (8)$$

For micrometer-scale displacement sensing around 300 GHz, this condition is well satisfied. For example, with $B = 5$GHz and $\Delta d = 10\ \mu$m, the phase variation across the sweep is only about 0.002 rad. When Eq. (8) holds, the displacement-induced phase increment becomes nearly constant over the swept bandwidth. If complex signals are available, this phase term can be directly estimated from their relative phase. The displacement $\Delta d$ can then be determined using an effective carrier frequency $f_c$ through

$$\Delta\phi_0(f) \approx \Delta\phi_0(f_c) = \frac{4\pi f_c}{c}\Delta d. \qquad (9)$$

This interferometric radar scheme is particularly suitable for the RTD platform, since RTDs provide a high carrier frequency and therefore enable micrometer-scale displacement sensitivity without requiring an extremely large sweep bandwidth. However, unlike the conventional radar case, a reliable complex signal is not directly available for self-mixing measurement because the proposed system does not include an I/Q demodulation chain. Therefore, $\Delta d$ cannot be directly estimated from Eqs. (7) and (9) in the same way as in a conventional radar system.

Instead, Eq. (9) also suggests a natural waveform-shift interpretation: an approximately constant phase offset can be represented as a small shift along the frequency axis [23,25]. If the measured sweep is regarded as a periodic waveform driven by a monotonically varying phase variable, then the displaced signal can be approximated as [25]

$$S_2(f) \approx S_1(f + \Delta f), \qquad (10)$$

where $\Delta f$ is the equivalent frequency shift associated with $\Delta d$. Since the phase varies linearly with frequency under the narrowband approximation, $\Delta f$ satisfies

$$\Delta\phi_0(f) \approx \frac{d\phi_0}{df}\Delta f \approx \frac{4\pi R}{c}\Delta f, \qquad (11)$$

By combining Eqs. (9) and (11), the relation between the equivalent frequency shift and displacement is obtained as

$$\Delta d \approx \frac{R}{f_c}\Delta f. \qquad (12)$$

Therefore, instead of directly estimating displacement from the phase of a complex signal using Eq. (7), the displacement can be inferred by estimating the constant frequency shift $\Delta f$ through waveform alignment using cross-correlation algorithm.

Notably, this waveform-shift interpretation can also be extended to self-mixing signals, provided that two conditions are satisfied. First, the phase mapping $\phi(\phi_0)$ must remain locally smooth and single-valued over the operating range, which is typically satisfied under weak-to-moderate feedback conditions without branch switching, as discussed in Sec. II(A). Second, the narrowband condition in Eq. (8) should remain valid, so that the displacement-induced increment in the free-running phase $\phi_0$ stays approximately constant over the frequency bandwidth. Under these conditions, a small increment in $\phi_0$ propagates to a corresponding increment in the feedback-modified phase $\phi$, and the shift model in Eq. (10) therefore remains valid even when the self-mixing waveform is distorted. Figure 4 provides a numerical example of this behavior: the simulated self-mixing waveforms for displacements up to 200 $\mu$m can still be approximated as shifted versions of the same underlying waveform for a center frequency of 285 GHz and a sweep bandwidth of 2 GHz.

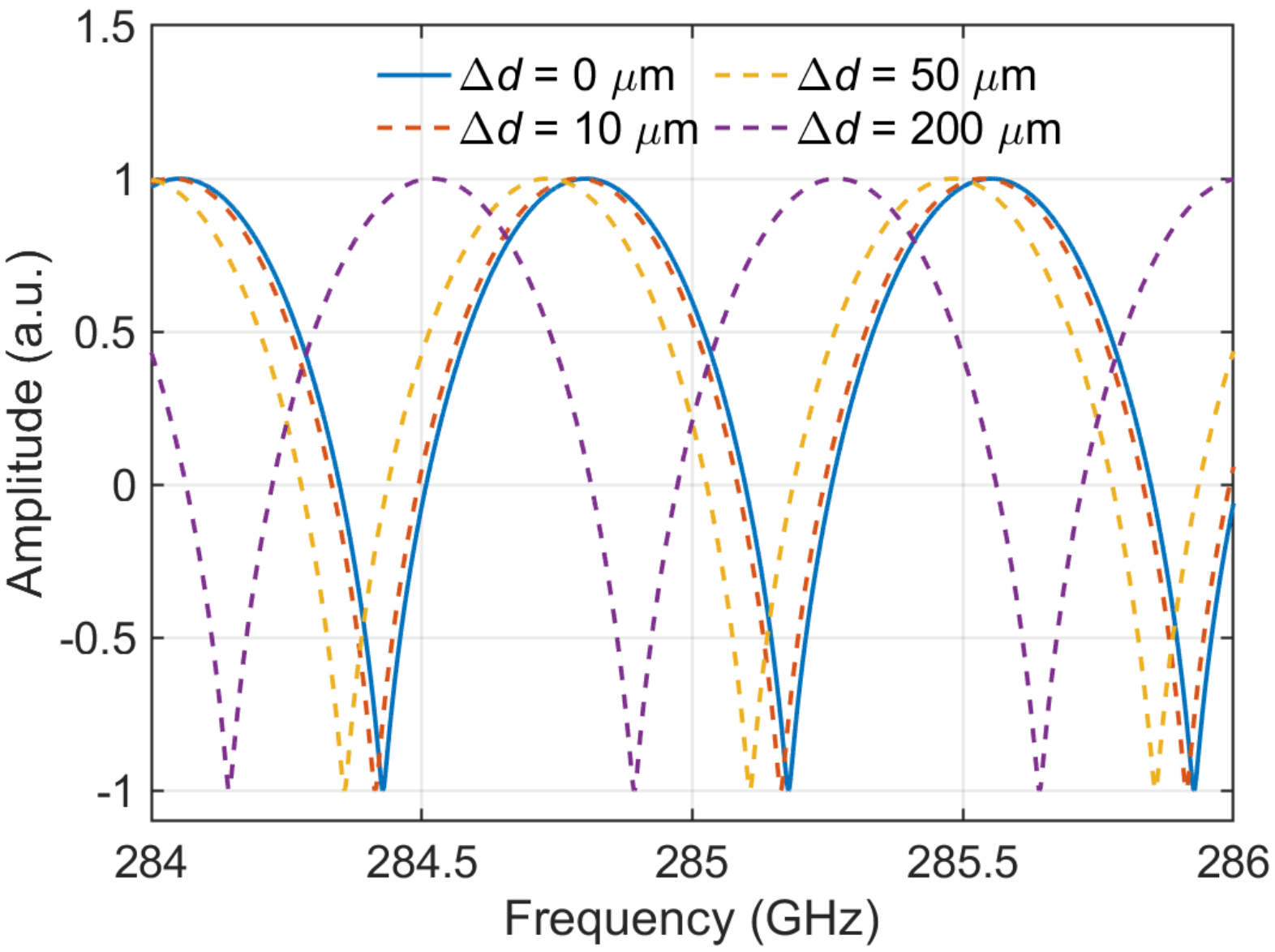

Fig. 4. Simulated self-mixing waveforms for different target displacements ($\Delta d$ = 0, 10, 50, and 200 μm) over a 2-GHz sweep, illustrating the approximate waveform shift used for displacement estimation (with $\boldsymbol{C} = \mathbf{0.7}$ and $\boldsymbol{\psi} = \mathbf{0}$).

## D. Phase-delay estimation for interferometric radar configuration

To robustly estimate small waveform shifts between the distorted self-mixing signals, we adopt a bounded normalized cross-correlation (NCC) approach [26,27]. After removing the mean from both signals, we computed the NCC within a prescribed lag window and selected the integer-sample delay as the lag that maximized the correlation. For each candidate integer shift $k$ within a predefined bound $k_{\text{min}} \leq k \leq k_{\text{max}}$, fixed-length overlapping segments of $S_1[n]$ and $S_2[n-k]$ are extracted. The mean value of each segment is removed prior to correlation, and the normalized cross-correlation coefficient is computed as

$$\rho(k) = \frac{\sum_{n\in\Omega_k}(S_1[n]-\bar{S}_{1k})(S_2[n-k]-\bar{S}_{2k})}{\sqrt{\sum_{n\in\Omega_k}(S_1[n]-\bar{S}_{1k})^2 \sum_{n\in\Omega_k}(S_2[n-k]-\bar{S}_{2k})^2}}, \qquad (13)$$

where $\Omega_k$ denotes the fixed-length overlap window and $\bar{S}_{1k}$, $\bar{S}_{2k}$ are the local means. The delay estimate $\hat{k}$ is obtained by maximizing $\rho(k)$ within the bounded interval. To achieve sub-sample accuracy, the peak location is further refined by three-point parabolic interpolation around the discrete maximum, using the peak and its two neighboring samples, which is a widely used strategy in correlation-based delay estimation [27]. The resulting estimate $\hat{k}$ represents the waveform shift in units of samples.

If the waveform shift is estimated by normalized cross-correlation as a delay of $\hat{k}$ samples, and the frequency increment per sample is $f_{\text{step}}$, then equivalent frequency shift $\Delta f = \hat{k} f_{\text{step}}$. Therefore, the displacement can be estimated using Eq. (12) as

$$\Delta d = \frac{R}{f_c}\hat{k} f_{\text{step}}. \qquad (14)$$

The bounded NCC estimator leverages the time shift between the two signals by correlating the entire waveform, thereby improving robustness through global information aggregation. To avoid the multi-peak ambiguity inherent in periodic self-mixing signals, the delay search is restricted to a bounded lag window smaller than half of the self-mixing period, thereby ensuring a unique correlation peak. This constraint, however, limits the unambiguous measurable displacement to within half a wavelength.

Based on the self-mixing model and the bounded NCC estimator, the numerical simulation results are summarized in Table I. For each condition, 500 repeated trials were performed with different displacement shifts, where $\sigma$ denotes the standard deviation of additive white Gaussian noise relative to the normalized signal amplitude. The root-mean-square error (RMSE) of the estimated shift was then recorded. The results suggest that the displacement estimation can achieve a high resolution of approximately 1 μm. Also, across a wide range of noise levels, the self-mixing cases ($C$ = 0.5 and $C$ = 0.9) achieve comparable or lower RMSE than the ideal sinusoid ($C$ = 0). This supports the intuition that self-mixing redistributes phase sensitivity across multiple high-slope intervals, so more samples carry useful displacement information.

Meanwhile, it should also be noted that the RMSE in Table I is approximately proportional to $\sigma$, which is physically reasonable because $\sigma$ directly represents the amplitude of additive noise in the simulation. Since the displacement is estimated through waveform alignment using normalized cross-correlation, the estimation error is expected to increase with the noise level under the present small-shift.

By contrast, the dependence of the RMSE on $\Delta d$ in Table I should not be interpreted as a fundamental trend under fixed SNR. In principle, for a valid local-shift model, the estimation error is not expected to depend strongly on $\Delta d$. In the present simulation, however, a larger $\Delta d$ makes the narrowband approximation in Eq. (8) less accurate, so the displaced waveform deviates slightly from an ideal shifted replica of the reference waveform. As a result, the estimation error may vary with $\Delta d$ in a weak and non-monotonic manner. Moreover, because the errors in Table I are obtained under highly ideal numerical conditions and are mostly below 1 $\mu$m, they are also sensitive to sampling density and discrete delay-estimation effects.

Table I. RMSE of the estimated displacement $\boldsymbol{\Delta d}$ under different additive noise levels (σ) and feedback factors $\boldsymbol{C}$. Values are given in μm.

| *Noise σ* | $\Delta d = 10\ \mu m$ | | | $\Delta d = 50\ \mu m$ | | | $\Delta d = 200\ \mu m$ | | |
|---|---|---|---|---|---|---|---|---|---|
| | $C$ = 0 | $C$ = 0.5 | $C$ = 0.9 | $C$ = 0 | $C$ = 0.5 | $C$ = 0.9 | $C$ = 0 | $C$ = 0.5 | $C$ = 0.9 |
| *0.01* | 0.089 | 0.098 | 0.074 | 0.163 | 0.182 | 0.132 | 0.083 | 0.099 | 0.068 |
| *0.05* | 0.590 | 0.621 | 0.491 | 0.614 | 0.675 | 0.496 | 0.583 | 0.622 | 0.534 |
| *0.1* | 1.191 | 1.001 | 0.913 | 1.104 | 1.103 | 0.993 | 1.616 | 1.323 | 1.000 |

## III. SYSTEM IMPLEMENTATION AND EXPERIMENT

### A. Integrated THz radar module using a single RTD

In this work, a packaged RTD module is used to demonstrate sensing in the 280-GHz band. The RTD chip has an area of approximately 0.025 $mm^2$ and is mounted behind a dielectric antenna, as shown in Fig. 1(b) [15]. The I–V characteristic is measured using a DC source, and the corresponding oscillation frequency is characterized with a spectrum analyzer, as shown in Fig. 5(a). Notably, to better satisfy the narrowband assumption for detecting smaller displacements, we choose a bias range of 0.36–0.365 V to ensure good frequency linearity, which provides an effective tuning span of approximately 2 GHz and satisfies the narrowband assumption, as shown in Fig.5(b). The output power of a single RTD device is approximately 20 μW.

The measurement configuration is summarized in Fig. 6(a). To acquire the RTD self-mixing signal, a low-frequency sawtooth bias waveform is applied to periodically sweep the device through the self-mixing region, thereby tuning the oscillation frequency over the selected range. The bias voltage $V_{\text{bias}}$ is generated by the digital-to-analog converter of a commercial field programmable gate array (FPGA) and connected to a T-connector (Port 1).

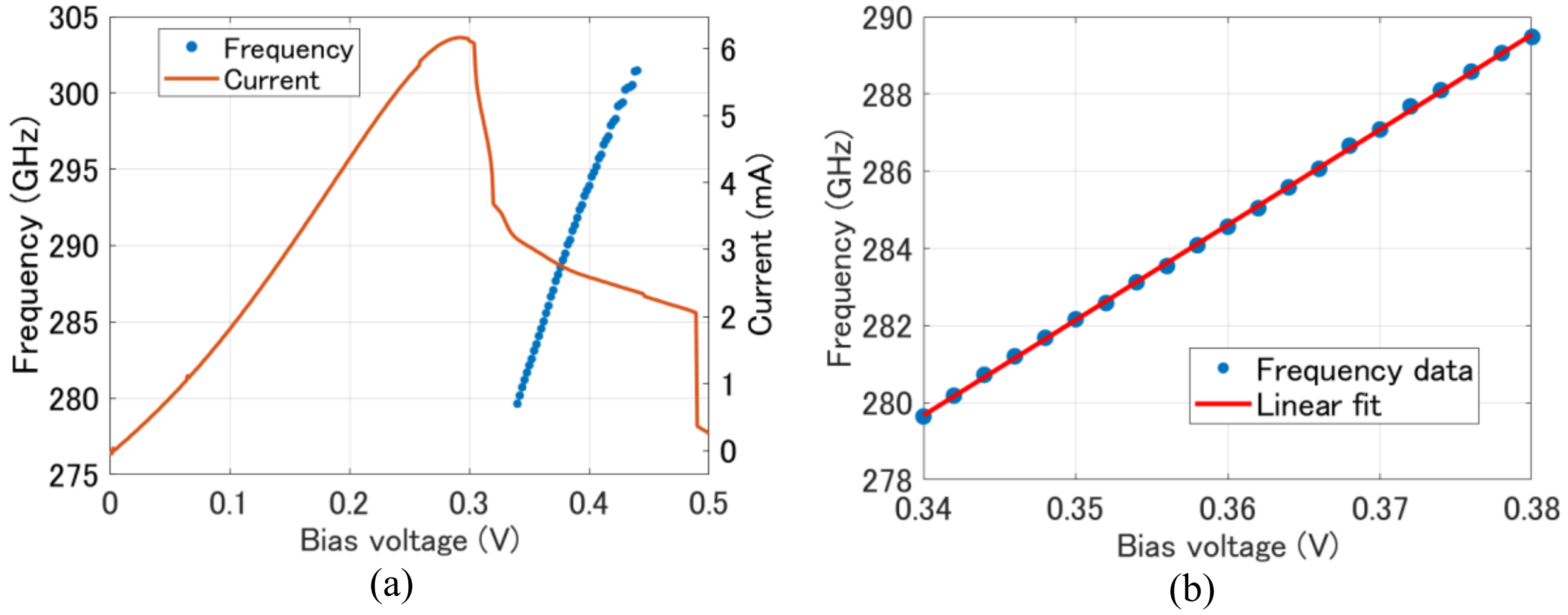

Fig. 5. (a) The corresponding oscillation frequency (left axis) of the packaged RTD module, and the RTD current–voltage (I–V) characteristic (right axis) of the packaged RTD module, with the effect of the shunt resistance compensated. (b) linear frequency tuning in the selected bias range (0.34–0.38 V) used for frequency sweeps.

During each sweep, the RTD connected to Port 2 of the T-connector operates in the self-mixing regime. The reinjected field induces a small self-mixing perturbation $V_{\text{RTD}}$, which is superimposed on the applied bias voltage $V_{\text{bias}}$. Since the self-mixing signal $V_{\text{RTD}}$ is much smaller than the applied bias, it is not directly observable at the RTD terminals. To extract the self-mixing signal, we employ a differential amplifier at Port3 of the T-connector to suppress the large bias component by injecting a synchronized bias voltage $V'_{\text{bias}}$. Finally, the amplified self-mixing signal $GV_{\text{RTD}}$is acquired using the analog-to-digital converter (ADC) of the same FPGA or an oscilloscope.

Because both the bias signal and the self-mixing signal reside in the baseband (typically kHz–MHz), the required electronics can be implemented with a standard arbitrary waveform generator and a moderate-speed digitizer. As shown in Fig. 6(b), an integrated RTD sensing system was developed using a commercial USB instrument together with a compact control circuit to generate the sawtooth bias and digitize the RTD self-mixing signal. This is in contrast to many THz coherent ranging architectures that require high-speed intermediate-frequency processing or an external coherent local oscillator.

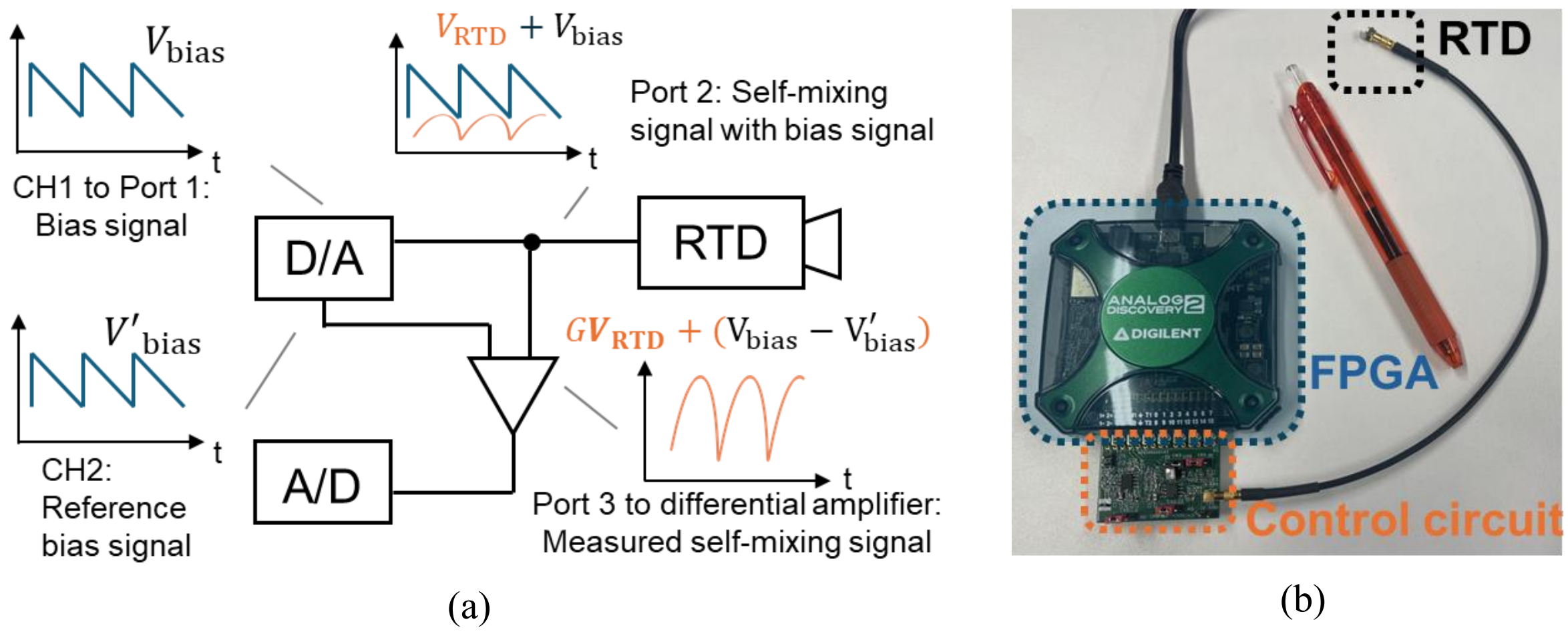

Fig. 6. (a) The configuration of the proposed RTD sensing system, (b) Photograph of the integrated prototype of the proposed system.

## B. Evaluation of the RTD sensing system

To evaluate the performance of the proposed system, self-mixing signals from a metallic reflector were measured at different distances using the experimental arrangement shown in Fig. 7(a). Figure 7(b) shows the sampled self-mixing signals for a metallic reflector placed at distances of 25, 40, and 70 cm, demonstrating the ranging capability and potential for three-dimensional imaging [18]. Owing to the high sensitivity of self-mixing signal [14,17], the radar signal remains detectable up to 70 cm even with a relatively low output power of approximately 20 μW.

Notably, although the THz beam was collimated at the source, beam divergence over longer propagation distances reduced the reflected power incident on the RTD. By fitting the measured self-mixing signal in Fig. 8 to the self-mixing model with a constant phase offset $\psi = 0.7$ and noise $\sigma = 0.05$, the effective feedback factor $C$ is estimated to be approximately 0.7 under the collimated condition. Notably, this value is expected to decrease when the reflector size is reduced or when the alignment is imperfect, since the returned power coupled back into the RTD becomes weaker.

The ADC captured the raw waveform directly; a simple moving average was applied to reduce system noise. For a metallic reflector placed 25 cm from the sensor, the relationship between modulation speed and the required averaging factor is summarized in Fig. 9. The standard deviation (STD) over 500 repeated measurements was computed from the phase offset estimated for each sweep using the method described in Sec. II(D). The mean value and corresponding STD quantify the stability of the self-mixing signal. As the modulation speed increases, a longer moving-average window is required to achieve comparable noise suppression. This may be related to reduced injection-locking stability when the transmitted and reflected signals change rapidly, leading to less effective phase-noise suppression [28]. In the following experiments, a modulation speed of 10 kHz was selected, and 1024 sweeps were averaged, yielding an acquisition time of approximately 10 ms. Each self-mixing sweep spanning 284–286 GHz was sampled with 1000 points, resulting in a frequency step $\Delta f$ of approximately 2 MHz.

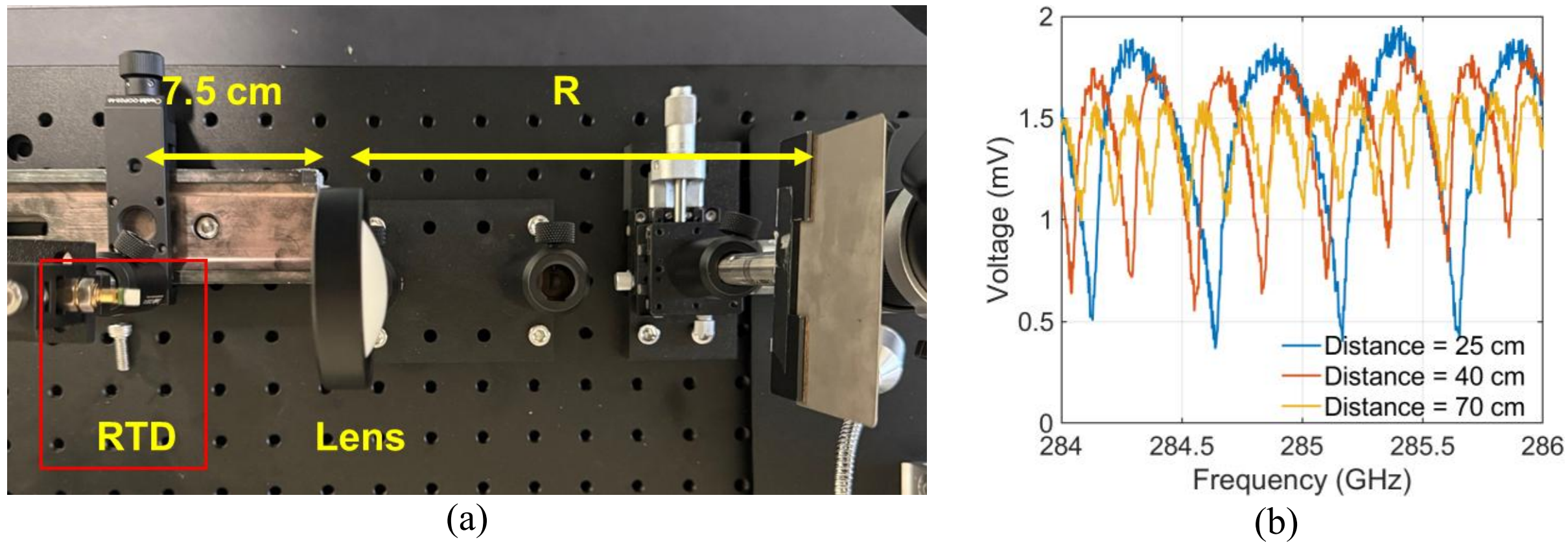

Fig. 7. (a) photograph of the experimental arrangement for ranging, and displacement measurements; (b) the obtained self-mixing signal of metallic reflector at different distances.

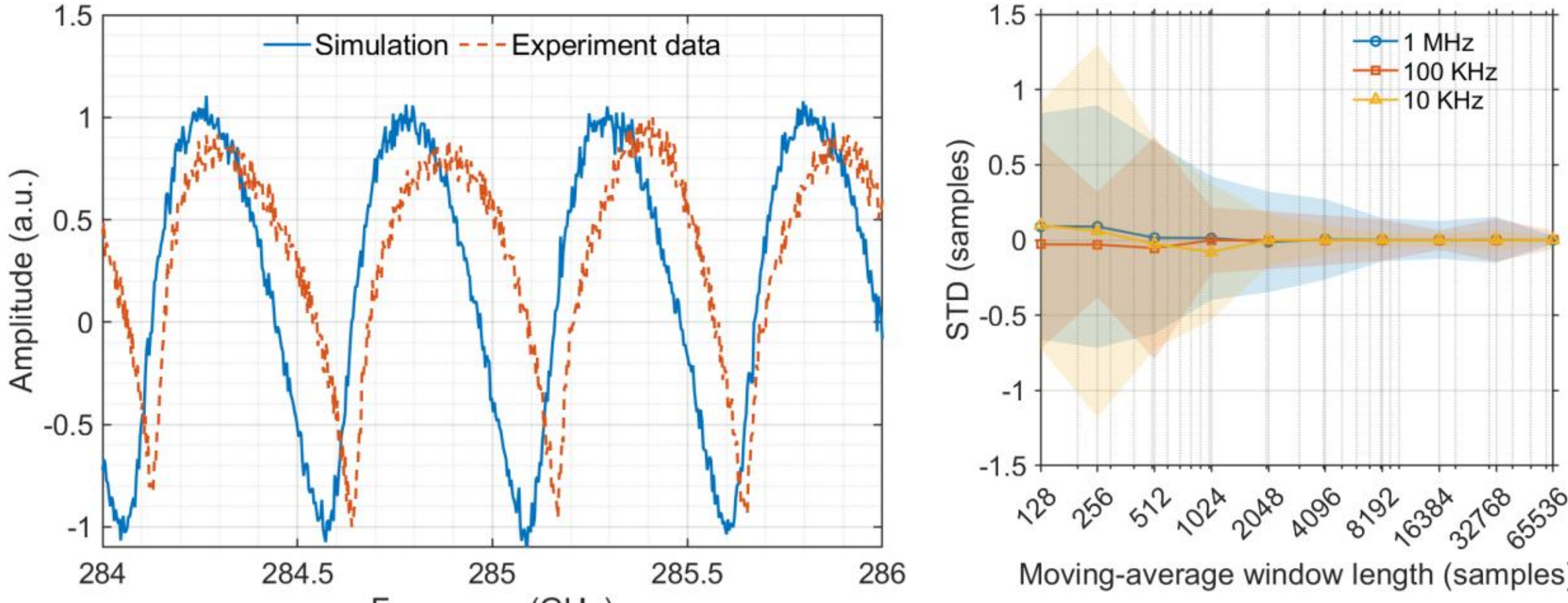

Fig.8 Simulated and experimentally measured self-mixing signal for a metallic reflector at 25 cm (with $\boldsymbol{C = 0.7}$, $\boldsymbol{\psi = 0.7}$ **and noise** $\boldsymbol{\sigma = 0.05}$).

Fig. 9 Relationship between sweep modulation speed and the required moving-average window length.

Finally, since the self-mixing signal is recorded versus bias voltage rather than versus oscillation frequency, the relation between the bias voltage and the oscillation frequency should be characterized, as shown in Fig. 5(b). However, measuring the RTD oscillation frequency requires a high-end spectrum analyzer, and the measured frequency can also vary with reflected power. In practice, a calibration can be obtained by measuring a reference target at a few known displacements and calibrating the system for a specific working distance using Eq. (10), which directly maps the bias-voltage shift to the displacement $\Delta d$. With this mapping, the displacement can be calculated directly using Eq. (12).

## C. Displacement and thin film thickness measurement

To demonstrate small-displacement sensing, the same experimental configuration as in Fig. 7(a) was used. As discussed in Sec. II(C), an interferometric radar approach was employed to retrieve small displacements beyond the bandwidth-limited range resolution. A metallic reflector positioned at 25 cm was used to acquire reference data. Controlled displacements were applied using a high-precision opto-mechanical translation stage to establish ground-truth values, and the displacement was estimated using the algorithm described in Sec. II(D). Figure 10 summarizes the displacement sensing results for commanded translation steps from 5 to 200 μm. For each commanded displacement, 50 repeated measurements were performed to evaluate repeatability. The estimated displacement agrees well with the ground-truth displacement, whereas the measurement spread increases for step sizes ranging from 10 μm to 200 μm, as shown in Fig. 10(a). A finer step size of 5 μm was also evaluated up to 50 μm, as shown in Fig. 10(b). The increased spread at smaller steps is consistent with the reduced phase excursion and the greater susceptibility to residual drift. The distributions of the measured displacements relative to the ground truth are presented as box plots. These results suggest that a practical displacement resolution on the order of ~10 μm is achievable with the proposed system under the current experimental conditions. The error is larger than the simulation results shown in Table I, even though Fig. 8 suggests a relatively low noise level in the self-mixing signal. This indicates that additive noise is not the dominant factor limiting the experimental accuracy, as discussed in Sec.II (d). Instead, the reduced accuracy, especially for small displacement steps, is more likely attributable to slow drift of the self-mixing waveform over time, temperature variations, and other experimental instabilities. In addition, because the small displacements were manually adjusted, the results may also include uncertainties associated with the translation stage and manual operation.

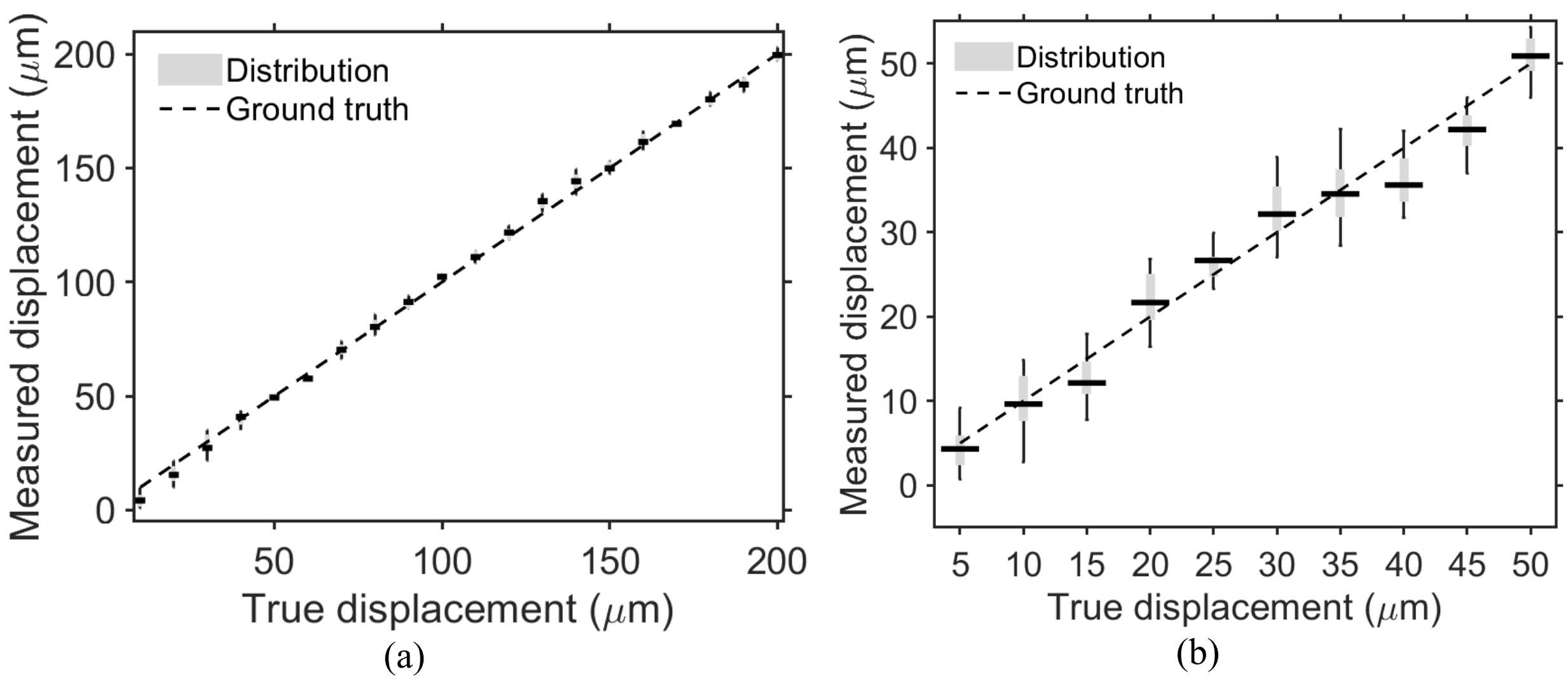

Fig. 10 Distribution of measured displacements at different true displacement values; (a) measurement results up to 200 μm, (b) measurement results up to 50 μm. Each box represents 50 repeated measurements, showing the median (central line), interquartile range (box), and whiskers (1.5× IQR). The dashed line indicates the ground truth.

Table II. Statistical summary of measured displacements ($N = 50$ per sample)

| Ground truth (μm) | Mean (μm) | Std. Dev. (μm) | Bias (μm) | RMSE (μm) |
| --- | --- | --- | --- | --- |
| 12.5 | 14.039 | 3.249 | 2.039 | 3.808 |
| 25.0 | 27.807 | 1.650 | 2.807 | 3.248 |
| 50.0 | 47.898 | 2.274 | −2.102 | 3.080 |

For thin-film thickness sensing, we consider a dielectric film with refractive index $n$ placed between the RTD and a metallic reference reflector, as shown in Fig. 11(a). Notably, an extra lens was included to obtain a better spatial resolution. Compared with the bare reflector, inserting a film of thickness change $\Delta z$ replaces an air path of $\Delta z$ with an optical path $n\Delta z$, resulting in an additional round-trip optical path difference of $2(n-1)\Delta z$. Accordingly, the film-induced change can be treated as an equivalent air-path displacement $(n-1)\Delta z$, leading to

$$\Delta z \approx \frac{kR\Delta f}{f_c(n-1)}. \qquad (15)$$

Table II. summarizes the thickness estimation results for polymer films with nominal thicknesses of 12.5, 25, and 50 μm. Each sample was measured repeatedly to evaluate the measurement statistics. An RMSE of approximately 3 μm is obtained across the three samples. As the target is translated laterally, the estimated thickness shows step-like transitions corresponding to the inserted film thickness. Notably, the refractive index of the sample was estimated to be approximately 1.7, which is in good agreement with the manufacturer-specified value. Since the equivalent air-path displacement associated with a 12.5 μm film is on the order of ~8 μm, this thickness is close to the practical detection limit of the proposed approach under the current experimental conditions.

For a more practical demonstration, the films were secured in a 3D-printed holder mounted on a translation stage for thickness-difference measurements. In this case, each lateral position was measured once using a focused THz beam, yielding an effective spatial resolution of approximately 2 mm to separate adjacent samples. Despite uncertainties in the effective refractive index and film positioning, the results demonstrate that micrometer-scale thickness steps can be resolved with the proposed system, as shown in Fig. 11(b).

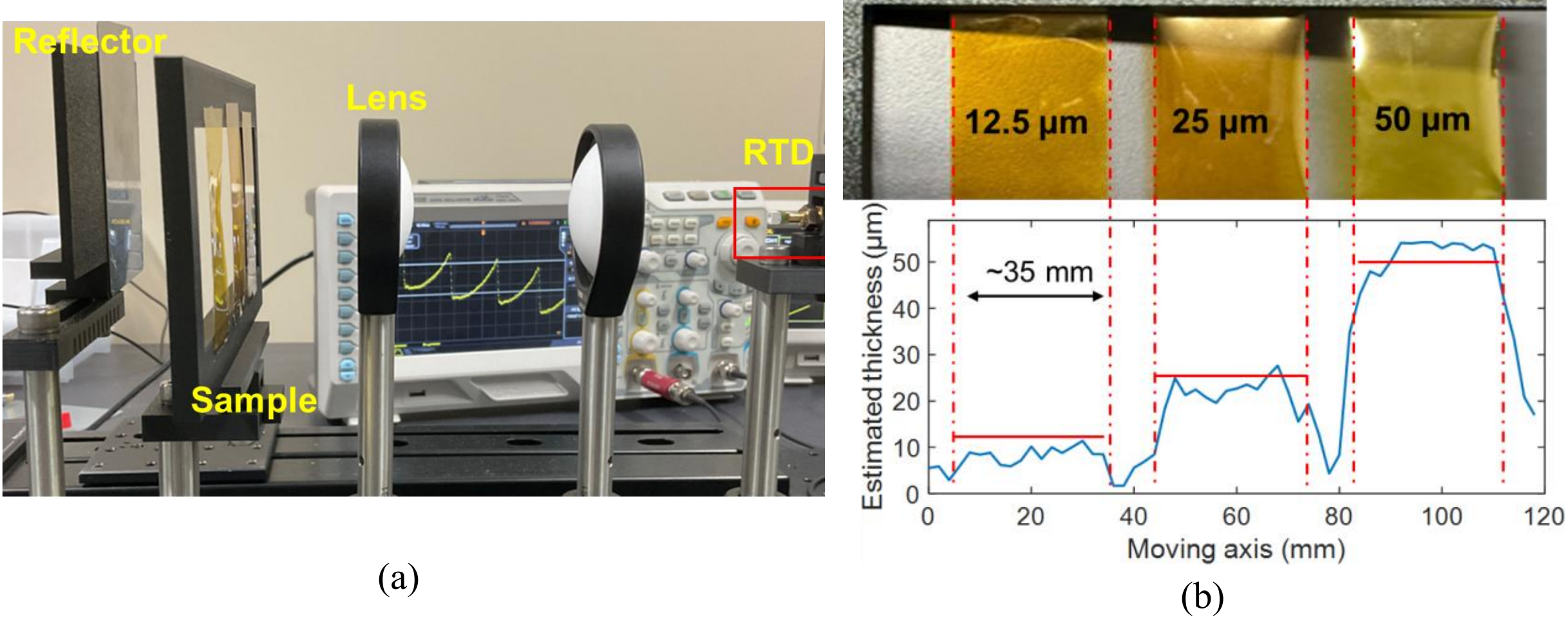

Fig. 11 (a) Photograph of the experimental setup; (b) estimated film thickness along a lateral scan over regions covered by 12.5 μm, 25 μm, and 50 μm polymer films.

## IV. DISCUSSION

The results highlight several practical features of RTD self-mixing metrology. First, the architecture is inherently compact because coherent mixing occurs directly inside the RTD, and only baseband electronics are required for bias control and signal readout. Second, self-mixing nonlinearity can be beneficial for small-shift estimation by providing multiple high-slope regions within a single interference period. However, strong external feedback can destabilize the oscillator and may necessitate the use of an external attenuator. In addition, it is important to clearly define the operating scenario to avoid large variations in the feedback factor $C$.

To fully exploit the advantages of the RTD sensor, an interferometric radar approach is introduced to leverage its high center frequency, making it well suited for monitoring small vibrations and detecting thickness variations in thin films. In these scenarios, the phase information at the center frequency becomes critical for accurate detection, enabling the possibility of resolving micrometer-scale displacements with the proposed approach. Moreover, since the displacement-to-phase conversion scales with frequency, operation at higher RTD oscillation frequencies (e.g., in the 2 THz band [14]) could, in principle, further improve the displacement sensitivity toward the nanometer scale, provided that the signal-to-noise ratio and phase stability are sufficient. On the other hand, small displacements or thin dielectric layers only weakly perturb the feedback condition, so the feedback factor $C$ does not vary abruptly, allowing the waveform-shift approximation described in Sec. II(C) to remain valid. However, the proposed method is inherently constrained by the characteristics of the RTD, which offers high operating frequency but limited tuning bandwidth due to the NDC regime. As indicated by Eq. (8), both increasing the bandwidth and the displacement magnitude can lead to the breakdown of the shift model, resulting in a trade-off between sweep bandwidth and model validity. Therefore, the proposed approach is primarily intended for inspection tasks involving small displacements within a fraction of the wavelength, such as thin-layer thickness estimation.

In the present thickness demonstration, a single effective refractive index is used to convert frequency shifts into thickness variations. For quantitative metrology of unknown materials, the refractive index may be obtained through independent characterization methods (e.g., THz-TDS) or by exploiting multi-frequency information across the frequency bandwidth. Although it is mathematically possible to invert thickness changes in multilayer structures, the resulting computational complexity limits the practicality of such approaches [10]. Consequently, the proposed method is more suitable for detecting thickness variations or localized defects in layered samples.

It should be noted that, owing to the unique operating mechanism of RTD self-mixing, it remains challenging to quantitatively characterize the phase noise and ultimate sensitivity of the proposed technique. In addition, phase noise, which may be caused by the noise of the DAC board, the injection power into the RTD, and temperature variations, limits the minimum achievable estimation accuracy and warrants further investigation from a device-physics perspective.

Finally, the proposed system operates entirely with low-frequency circuitry in the sub-MHz range, which facilitates straightforward extension to a monostatic array configuration using only microwave components. When combined with high-speed switching components, this architecture enables the realization of a THz-band imaging system in which each RTD simultaneously functions as a transmitter and a detector. Compared with THz array systems implemented using CMOS technology, the proposed configuration offers significant reductions in system size and power consumption.

## V. CONCLUSION

We presented a compact THz radar sensor based on a single resonant-tunneling diode operating near 280 GHz. By sweeping the RTD bias to generate a frequency sweep and measuring the resulting self-mixing waveform, micrometer-scale displacement and thin-film thickness changes can be inferred from small phase shifts using an interferometric radar approach. The proposed technique was validated using both numerical simulations and experimental measurements. The system demonstrated displacement sensing over 5–200 μm and resolved polymer film thickness steps down to 12.5 μm, highlighting the potential of single-device RTD self-mixing for compact THz metrology.

## Acknowledgement

This work was partially supported by the JST Core Research for Evolutional Science and Technology (CREST，JPMJCR21C4), and the Multidisciplinary Research Laboratory System of the Graduate School of Engineering Science, Osaka University.

## .REFERENCES